\documentclass[11pt, a4paper]{article}
\usepackage{jheppub}
\pdfoutput=1

\usepackage{amsmath, amsfonts, amssymb}

\usepackage{physics, yfonts, comment}

\newcommand{\sgn}{\mathrm{sgn}}
\newcommand{\heun}{H\ell}
\newcommand{\rs}{r_{b}}
\newcommand{\ws}{\textswab{w}}
\newcommand{\xx}{\widetilde{\rs}}

\newcommand{\rev}[1]{#1}
\newcommand{\revb}[1]{#1}

\title{Thermodynamics, magnetic properties, and global $U(1)$ symmetry breaking of the S-type Gubser-Rocha model}

\author{Shuta Ishigaki,}  
\author{Zhaojie Xu}
\affiliation{
Department of Physics, Shanghai University,
99 Shangda Road, Shanghai 200444, China}

\emailAdd{shutaishigaki@shu.edu.cn}
\emailAdd{zeezj@shu.edu.cn}

\abstract{
We study an explicit formula for the thermodynamic potential of the AdS dyonic black brane solution with an axio-dilaton hair, which was discovered in an extension of the $(3+1)$d Gubser-Rocha model enjoying S-duality.
From the thermodynamic potential, we can compute the magnetization and the magnetic susceptibilities of the dyonic solution.
The result of the magnetization is negative implying that the system is diamagnetic.
Subsequently, we consider a specific neutral limit of the dyonic solution.
In this setup, we find that the system exhibits spontaneous breaking of a global $U(1)$ symmetry.
The order parameter is given by a complex operator which is dual to the axio-dilaton field in the bulk.
Interestingly, the system has a finite Hall conductivity even in the absence of the external magnetic field, and it is related to the phase of the complex operator.
}
\makeatletter

\begin{document}

\maketitle

\section{Introduction}

The holographic method has been widely applied for studying strongly coupled condensed matter systems \cite{Hartnoll:2016apf}.
One of the advantages of holography is that it allows us to analyze strongly coupled many-body quantum systems non-perturbatively. A great deal of the boundary theory can often be understood by solving classical equations of motion of the bulk.

In a real experiment, magnetic fields are often utilized to investigate various properties of a material.
In holographic studies of $(2+1)$d boundary theories, the charged system in the presence of the out-of-plane external magnetic field can be described by the corresponding black brane solution with both the electric and the magnetic charges, i.e., the dyonic black brane.
Investigating various properties of the dyonic solutions of the holographic models is, therefore, useful to test the holographic models by comparing the magnetic properties with the real measurements.
It has been studied for various holographic models in both analytic and numerical ways as in~\cite{Hartnoll:2007ai,Goldstein:2010aw,Lu:2013ura,Chow:2013gba,
Kim:2015wba,Seo:2015pug,Zhou:2018ony,Kim:2019lxb,Ahn:2023ciq}.

Recently, a new analytic solution of a dyonic AdS black brane was found in a model with S-duality \cite{Ge:2023yom}.
The model can be thought of as a usual Einstein--Maxwell-Dilaton-Axion theory with a linear axion term for breaking the translational invariance.
In particular, the model is extended from the original $(3+1)$d Gubser-Rocha model \cite{Gubser:2009qt} by incorporating S-duality.
The extension might be natural because the Gubser-Rocha model can be lifted to $11$d supergravity/M-theory.
By virtue of the S-duality, one can find an analytic dyonic solution.
The dyonic solution has nontrivial dilaton and axion profiles, which is also called \emph{axio-dilaton} collectively.
Using the analytic dyonic solution, we can investigate its property including the nonlinear effects of the external magnetic field.
It is interesting to study several properties of this analytic dyonic solution for understanding its dual description in the boundary field theory.

In this paper, we investigate several aspects of this analytic dyonic solution.
In the former part, we compute the thermodynamic potential density of the axio-dilatonic dyonic solution.
In order to compute the thermodynamic potential density of the dyonic black brane solution with the axio-dilaton hair, we consider the deformation of a multi-trace operator \cite{Witten:2001ua}, and perform the holographic renormalization \cite{Skenderis:2002wp} following a similar way studied in~\cite{Caldarelli:2016nni}.
We obtain the expression of the thermodynamic potential, which is consistent with the previous studies for $B=0$.
By using the result of the thermodynamic potential, we compute the analytic expressions for magnetization and magnetic susceptibilities of the dyonic black brane.
We find that the magnetic susceptibility always takes negative values, so the system exhibits diamagnetism.
As a result of the multi-trace deformation, we can regard the leading term of the axio-dilaton field as the vacuum expectation value (vev) of the dual complex-valued operator, which has an analytic expression.

In the latter part, we consider a neutral limit while keeping the ratio between the electric and the magnetic charge fixed.
It leads to a neutral black brane spacetime with nontrivial axio-dilaton profile.
We find that the neutral solution exhibits a phase transition between the axio-dilaton black brane and the bald black brane solution.
\rev{%
We can understand this transition as spontaneous breaking of a global $U(1)$ symmetry in our choice of the triple-trace marginal deformation.
We also find the corresponding global symmetry in the bulk action.
Note that the global symmetry is not gauged, so the vector field $A_{\mu}$ does not represet the current operator associated with this global $U(1)$ symmetry.
}%
Remarkably, the phase of the complex operator $\xi$ is related to the components of the DC conductivity tensor.
In particular, we find that the system exhibits a finite Hall conductivity even in the absence of the external magnetic field and exhibits the spontaneous magnetization when the axion has a nontrivial profile.

We also analyze the AC conductivity in the neutral axio-dilatonic black brane.
The AC conductivity can be computed by the vector field perturbations.
In this case, the solutions for the vector perturbations are given by terms of the Heun's function, so we can also write the AC conductivity by using the Heun's function formally. Along the way, we also investigate the corresponding quasinormal modes.
We find that the quasinormal modes have a finite real-part of the frequency correspondingly, when the Hall conductivity becomes finite.

This paper is organized as follows.
In section \ref{sec:dyonic}, we briefly review our model and the dyonic solution found in ref.~\cite{Ge:2023yom}.
We compute the thermodynamic potential by considering multi-trace deformations.
By using the result, we also compute the magnetization and the magnetic susceptibilities of this solution.
In section \ref{sec:neutral}, we consider the neutral limit of the dyonic solution.
In this limit, we obtain the analytic formula of the AC conductivity.
Section \ref{sec:discussion} is devoted to the conclusions and discussions.
Some results of a phase diagram in the case of adding a charged scalar sector are provided in appendix \ref{appendix:charged_sector}.

\section{Properties of the dyonic S-completed Gubser-Rocha solution}\label{sec:dyonic}
In this study we consider the model introduced in ref.~\cite{Ge:2023yom}.
We briefly review the analytic dyonic solution of this model and some of its properties.
The model consists of two parts: the S-completed Gubser-Rocha model, and the translational symmetry breaking axion term.%
\footnote{
    In Appendix \ref{appendix:charged_sector}, we also consider a charged scalar sector in addition.
}
The first part is a S-completed version of the $(3+1)$d Gubser-Rocha model proposed in \cite{Ge:2023yom}, whose Lagrangian density is given by
\begin{equation}
	\frac{\mathcal{L}_1}{\sqrt{-g}}
	=
	R - \frac{3}{2}\frac{\partial_{\mu}\tau\partial^{\mu}\bar{\tau}}{(\Im\tau)^2}
	- \frac{1}{4}e^{-\phi} F^2 + \frac{1}{4} \chi F \tilde{F} + \frac{3}{L^2} \frac{\tau\bar{\tau} + 1}{\Im \tau},
\end{equation}
where $\tau = \chi + i e^{-\phi}$ is the axio-dilaton, and $\tilde{F}^{\mu\nu} = \frac{1}{2 \sqrt{-g}} \epsilon^{\mu\nu\rho\sigma} F_{\rho\sigma}$ where $L$ is a length scale corresponding to the AdS radius.
In terms of the axion $\chi$ and the dilaton $\phi$, we write
\begin{equation}\label{eq:S-completed}
	\frac{\mathcal{L}_1}{\sqrt{-g}}
	=
	R - \frac{3}{2} (\partial \phi)^2 - \frac{3}{2} e^{2\phi} (\partial\chi)^2 
	-\frac{1}{4} e^{-\phi} F^2 + \frac{1}{4} \chi F \tilde{F} + \frac{1}{L^2}(6\cosh\phi + 3\chi^2 e^{\phi}).
\end{equation}
This Lagrangian density is invariant under the following `S-dual' transformation.
\begin{equation}
	\tau \to  \tau' =- 1/\tau\quad
	F \to  F' = \chi F + e^{-\phi} \tilde{F},\quad
	\tilde{F} \to \tilde{F} ' = \chi \tilde{F} - e^{-\phi} F,
\end{equation}
so the model is S-complete.
Furthermore, we can show that the Lagrangian is actually invariant under a $SO(2)\simeq U(1)$ subgroup of $SL(2;\mathbb{R})$. The corresponding transformation on the axio-dilaton and the field strength is given by
\begin{align}
    \begin{split}
         \tau &\to  \tau' =\frac{\tau\cos\theta+\sin\theta}{-\tau\sin\theta+\cos\theta},\\
         F &\to  F' = (-\chi\sin\theta+\cos\theta) F -\sin\theta\, e^{-\phi} \tilde{F},\\
         \tilde{F} &\to \tilde{F} ' = (-\chi\sin\theta +\cos\theta) \tilde{F} +\sin\theta\, e^{-\phi} F,
    \end{split}
    \label{eq:U(1)_symmetry}
\end{align}
which plays a crucial role in our discussion for later sections.
Expanding the potential term for small $\phi$ and $\chi$, we read the masses of the axion and the dilaton as $m_{\chi}^2=m^2_{\phi} = - 2/L^2$.
From the well-known formula $\Delta(\Delta - d) = m^2 L^2$ for scalar fields in an asymptotic AdS$_{d+1}$ spacetime, we obtain the scaling dimensions as $\Delta_{-}=1$ and $\Delta_{+}=2$.
Remarkably, both masses take values in the range of $-d^4/4 \leq m^2 L^2 \leq -d^2/4 +1$, which implies we can consider multi-trace deformations as we will show later.
The second part is a linear axion model to break the translational symmetry by hand, whose Lagrangian density is given by
\begin{equation}
    \frac{\mathcal{L}_2}{\sqrt{-g}} = - \frac{1}{2}\sum_{I=1,2}(\partial \psi_{I})^2.
\end{equation}
The total action is thus given by
\begin{equation}
    S = \frac{1}{2\kappa^2}\int\dd[4]{x}\left(
        \mathcal{L}_{1} + \mathcal{L}_{2}
    \right),
\end{equation}
where $\kappa$ is the gravitational constant.
Hereafter, we set $2\kappa^2 = L = 1$ for simplicity.

In this system, the following dyonic black brane solution was analytically obtained in \cite{Ge:2023yom}.
For the metric ansatz of
\begin{equation}
	\dd{s}^2 = - f(r) \dd{t}^2 + \frac{\dd{r}^2}{f(r)} + h(r)(\dd{x}^2 + \dd{y}^2),
\end{equation}
the solution is obtained as
\begin{equation}\label{eq:solution_1}
\begin{gathered}
	f(r) = h(r)\left(
		1 - \frac{n^2 + B^2}{3 \rs(\rs + r)^3} - \frac{k^2 \rs}{2 \rs(\rs + r)^2}
	\right),\quad
	h(r) = \sqrt{r(r+\rs)^3},\\
	e^{-\phi} = \frac{
		(n^2 + B^2)\sqrt{r(r+\rs)}
	}{
		(n^2 + B^2)r + B^2 \rs
	},\quad
	\chi = - \frac{B n \rs}{(n^2 + B^2) r + B^2 \rs},\\
	A = n \left(\frac{1}{r_0 + \rs} - \frac{1}{r + \rs}\right)\dd{t} - \frac{B}{2}y \dd{x} + \frac{B}{2}x \dd{y},\quad \psi_{I} = (k x, k y),
\end{gathered}
\end{equation}
where $B$ is an external magnetic field which is orthogonal to the $x$--$y$ plane.
$k$ is a strength of the momentum relaxation.
An outer horizon is located at $r=r_0$, while the AdS boundary is located at $r=\infty$.
$\rs$ is a real parameter determining the location of the curvature singularity at $r=-\rs$.
$n$ is related to the other parameters by
\begin{equation}\label{eq:density}
	n = \sqrt{3\rs(r_0 + \rs)^3\left(
		1 - \frac{B^2}{3\rs(r_0 + \rs)^3} - \frac{k^2}{2(r_0 + \rs)^2}
	\right)}.
\end{equation}
It corresponds to the charge density in the boundary theory, and the chemical potential $\mu$ is given by $\mu = n/(r_0 + \rs)$.
The Hawking temperature and entropy density are obtained as
\begin{equation}
	T = \frac{r_0}{8\pi\sqrt{r_0(r_0 + \rs)^3}}\left(
		6(r_0+\rs)^2 - k^2
	\right),\quad
	s = 4\pi\sqrt{r_0(r_0 + \rs)^3},
\end{equation}
respectively.
The family of solution is parameterized by $(r_0, \rs, k, B)$ corresponding to the boundary quantities $(T, \mu, k, B)$.

\rev{
We remark that the transformation (\ref{eq:U(1)_symmetry}) acts on the solution (\ref{eq:solution_1}) as a rotation of the charge density and the magnetic field, which is written as
\begin{equation}
    \begin{pmatrix}
        n \\ B
    \end{pmatrix}
    \to
    \begin{pmatrix}
        n' \\ B'
    \end{pmatrix}
        =
    \begin{pmatrix}
        \cos\theta & -\sin\theta\\
        \sin\theta & \cos\theta
    \end{pmatrix}
    \begin{pmatrix}
        n \\ B
    \end{pmatrix}.
\end{equation}
}%
\revb{
It implies that the new solution corresponds to a solution with different boundary conditions, namely ($n'$,$B'$).
}

With this solution (\ref{eq:solution_1}), the components of the DC conductivity tensor are obtained as \cite{Ge:2023yom}
\begin{equation}\label{eq:DC_conductivities}
\begin{gathered}
    \sigma_{xx} = \sigma_{yy}
    =
    \left.
    \frac{
      k^2 h(r) \left(
        B^2 + k^2 h(r)e^{\phi} + h(r)^2 F_{rt}^2
      \right)
    }{
        (B^2+k^2e^{\phi}h(r))^2 + B^2 h(r)^2 F_{rt}^2
    }
    \right|_{r=r_{0}},\\
    \sigma_{xy}
    = - \sigma_{yx}
    =
    \left.
    \frac{
        B h(r) e^{-\phi} F_{rt} (2k^2 h(r) e^{\phi} + h(r)^2 F_{rt}^2 +B^2)
    }{
        (B^2 + k^2 e^{\phi} h(r))^2 + B^2 h(r)^2 F_{rt}^2
    } - \chi
    \right|_{r=r_0}.
\end{gathered}
\end{equation}
The components of the thermoelectric conductivity tensor are also given by
\begin{equation}\label{eq:DC_thermoelectric_conductivities}
\begin{gathered}
    \alpha_{xx} = \alpha_{yy} =
    \left.
    \frac{
        4\pi k^2 h(r)^3 e^{\phi} F_{rt}
    }{
        (B^2 + k^2 e^{\phi} h(r))^2 + B^2 h(r)^2 F_{rt}^2
    }\right|_{r=r_0},\\
    \alpha_{xy} = - \alpha_{yx}
    =
    \left.
    \frac{
        4\pi B h(r) (B^2 + k^2 e^{\phi} h(r) + h(r)^2 F_{rt}^2)
    }{
        (B^2 + k^2 e^{\phi} h(r))^2 + B^2 h(r)^2 F_{rt}^2
    }\right|_{r=r_0}.
\end{gathered}
\end{equation}
The Hall angle is given by
\newcommand{\PP}{P}
\begin{equation}\label{eq:Hall_angle}
    \tan\theta_{H} = \frac{\sigma_{xy}}{\sigma_{xx}}
    =
    \frac{\mu \PP\left(
        2k^2 (\mu^2+\PP^2)(r_0+\rs)
        + k^4 \rs
        + (\mu^2 + \PP^2)^2(r_0 + \rs)
    \right)}{
        k^2(\mu^2 + \PP^2)
        (k^2 + \mu^2 + \PP^2)
        \sqrt{r_0(r_0 +\rs)}
    },
\end{equation}
where $\PP=\frac{B}{r_0 + \rs}$.

\subsection{Thermodynamic potentials}
The goal of this subsection is to determine the thermodynamic potential of the dyonic solution,  following a similar procedure as in ref.~\cite{Caldarelli:2016nni}.
Substituting the solution into the Lagrangian density, we obtain
\begin{equation}
	\mathcal{L}_{1} = k^2 + \frac{n^2 - B^2}{2 (r+\rs)^2} - 3 (r+\rs)(2 r + \rs),\quad
	\mathcal{L}_{2} = - k^2.
\end{equation}
As usual holographic models, the onshell action involves divergences, so we have to perform the holographic renormalization.

In order to perform the holographic renormalization, we have to fix the coordinate gauge to the so-called Fefferman-Graham (FG) coordinate of the form
\begin{equation}
    \dd{s}^2 = \frac{h_{ij}(z)\dd{x^i}\dd{x^j} + \dd{z}^2}{z^2},
\end{equation}
where $z$ denotes the FG coordinate, and $i,j = t,x,y$.
The coordinate transformation can be obtained by solving $\dd{z}^2/z^2 = \dd{r}^2/f(r)$ in a series expansion near the boundary, which is given by
\begin{equation}
	r = \frac{1}{z} - \frac{3}{4}\rs + \frac{1}{64}(8k^2 + 3\rs^2) z +
	\left(
		\frac{n^2 + B^2}{18 \rs} - \frac{k^2 \rs}{24} + \frac{\rs^3}{96}
	\right) z^2 + \order{z^3}.
\end{equation}
Using this, the metric components are expanded as
\begin{equation}\label{eq:FG_expansions_metric}
\begin{aligned}
	-h_{tt} =& 1  + \frac{ - 8k^2 - 3 \rs^2}{32} z^2
	+ \left(
		- \frac{2}{9} \frac{B^2 + n^2}{\rs} + \frac{k^2 \rs}{6} - \frac{\rs^3}{24}
	\right) z^3
	+ \order{z^4},\\
	h_{xx} =& 1 + \frac{8 k^2 - 3\rs^2}{32} z^2
	+\frac{8 (B^2 + n^2) - 6 k^2 \rs^2 - 3 \rs^4}{72\rs} z^3
	+ \order{z^4},
\end{aligned}
\end{equation}
The dilaton and the axion are expanded as
\begin{equation}\label{eq:FG_expansions_axiodilaton}
\begin{aligned}
	\phi = & \frac{\rs}{2} \frac{B^2 - n^2}{B^2 + n^2} z
	+ \frac{(B^4 + 4 B^2 n^2 - n^4)\rs^2}{8(B^2 + n^2)^2} z^2 + \order{z^3},\\
	\chi =& - \frac{B n \rs}{B^2 + n^2} z
	+
	\frac{B n (B^2 - 3 n^2) \rs^2}{4 (B^2 + n^2)^2} z^2+ \order{z^3}.
\end{aligned}
\end{equation}
We also write the coefficients of the FG expansions as $z^{-1}\phi = \phi_{(0)} + \phi_{(1)}z + \order{z^2}$, and for $\chi$ in the same manner, for later use.
We regularize the onshell action by introducing a cutoff slice at $z=\epsilon$ for small positive $\epsilon$:
\begin{equation}
    S_{\text{bulk}} :=
    \int\dd[3]{x}
    \int_{r_0}^{r(z=\epsilon)}\dd{r}\left(
        \mathcal{L}_{1} + \mathcal{L}_{2}
    \right).
\end{equation}
We have to also consider the Gibbons-Hawking term at the same slice:
\begin{equation}
    S_{\text{GH}} :=
    2\int_{z=\epsilon}\dd[3]{x} \sqrt{-\gamma} \gamma^{ij} K_{ij},
\end{equation}
where $K_{ij} = - \frac{1}{2}z \partial_{z} \gamma_{ij}$ is an extrinsic curvature, and $\gamma_{ij}$ is an induced metric at the slice $z=\epsilon$.

In the standard procedure of the holographic renormalization, the renormalized action is given by
\begin{equation}
    S_{\text{ren}} := \lim_{\epsilon \to 0} \left[
        S_{\text{reg}}
        + S_{\text{ct}}
    \right],
\end{equation}
where $S_{\text{reg}}=S_{\text{bulk}} + S_{\text{GH}}$, and $S_{\text{ct}}$ is a counterterm we take
\begin{equation}
\begin{aligned}
	S_{\text{ct}} :=
    - \int\dd[3]{x}
    \int_{z=\epsilon} \sqrt{-\gamma}\left[
		4 + R[\gamma]
        - \frac{1}{2}\sum_{I} \gamma^{ij}\partial_{i}\psi_{I}\partial_{j}\psi_{J}
		+ \frac{3}{2} \phi^2 + \frac{3}{2} e^{\phi} \chi^2
	\right],
\end{aligned}
\end{equation}
where $R[\gamma]$ denotes the scalar curvature of the induced metric.
Note that the boundary geometry with $\gamma_{ij}$ is flat in our case.
By using the solution, we obtain
\begin{equation}
    S_{\text{ren}} =
    \int\dd[3]{x}\left[
        \frac{\rs^3}{16} + (r_0 + \rs)^3 + \frac{k^2 r_0}{2} - \frac{B^2}{r_0 + \rs}
    \right].
\end{equation}
However, this is a renormalized action in standard quantization.
In the standard quantization, the leading terms of the dilaton and the axion in the vicinity of the AdS boundary are understood as external sources to the corresponding operators.
By performing variation, we obtain the vevs of the dual operators as
\begin{equation}
    \expval{O_{\phi}}_{\rm std}=3\left(\phi_{(1)}-\frac{1}{2}\chi_{(0)}^2\right),\quad
    \expval{O_{\chi}}_{\rm std}=
    3\left(\chi_{(1)}+\phi_{(0)}\chi_{(0)}\right),
\end{equation}
where the bracket with $\rm std$ denotes the vev of the dimension $\Delta_{+}=2$ operators in the standard quantization.

In~\cite{Caldarelli:2016nni}, they showed that the black brane solution with a dilaton hair can be understood as a vacuum state without any supporting external source, by considering multi-trace operator deformations.%
\footnote{
    The authors of \cite{Chagnet:2022yhs} numerically found more general solutions in the original Gubser-Rocha model, corresponding to the cases with the finite source in several ways of the possible quantization.
    Our model should have such a numerical solution too.
}
The multi-trace deformation can be realized by introducing the appropriate finite boundary action, which corresponds to imposing a Robin boundary condition in the bulk picture \cite{Witten:2001ua}.
Since the masses of the dilaton and the axion take values within the range for the multi-trace deformations, we can apply a similar prescription in our case, but we have to extend the prescription for the single scalar hair to the axio-dilaton hair.
In this study, we consider the finite boundary action given by
\begin{equation}\label{eq:Sfin}
\begin{gathered}
	S_{\text{fin}} := \int\dd[3]{x}\sqrt{-h_{(0)}}\left[
		\phi_{(0)} J_{\phi} + \chi_{(0)} J_{\chi} + \mathcal{F}(\phi_{(0)},\chi_{(0)})
	\right],\\
	J_{\phi} = - 3\left(\phi_{(1)} - \frac{1}{2}\chi_{(0)}^2\right)
	- \pdv{\mathcal{F}}{\phi_{(0)}},\quad
	J_{\chi} = - 3\left(\chi_{(1)} + \phi_{(0)} \chi_{(0)}\right)
	- \pdv{\mathcal{F}}{\chi_{(0)}},
\end{gathered}
\end{equation}
where $\mathcal{F}$ is a function of $\phi_{(0)}$ and $\chi_{(0)}$ which will be determined.
$J_{\phi}$ and $J_{\chi}$ are new source terms in the alternative quantization.
We take
\begin{equation}\label{eq:deformation_F}
	\mathcal{F} = - \sgn(\rs) \frac{1}{2}\left(\phi_{(0)}^2 + \chi_{(0)}^2\right)^{3/2},
\end{equation}
then the solution satisfies $J_{\phi} = J_{\chi} = 0$.
$\phi_{(0)}$ and $\chi_{(0)}$ are understood as vevs of dimension $\Delta_{-}=1$ operators.
Writing $\sigma_{(0)} = \phi_{(0)} + i \chi_{(0)}$, we obtain
\begin{equation}
    \mathcal{F} = -\sgn(\rs) \frac{1}{2}\abs{\sigma_{(0)}}^3.
\end{equation}
Therefore, this deformation can be understood as a triple-trace deformation of a complex-valued operator $O_{\sigma}$ whose vev is given by $\sigma_{(0)}$.
Note that since the deformation is given by the cubic term of the absolute value of $\sigma_{(0)}$, the deformed potential might be bounded below.
In addition, we can ignore the factor of $\sgn(\rs)$ in this study because the branch of $\rs>0$ corresponds to the ground states.

Substituting the expansions of the solution into eq.~(\ref{eq:Sfin}), we obtain
\begin{equation}
    S_{\text{fin}} = - \int\dd[3]{x}\frac{\rs^3}{16}.
\end{equation}
Finally, we obtain the renormalized onshell action in the alternative quantization as
\begin{equation}
    S_{\text{ren}}^{\text{alt}} :=
    S_{\text{ren}} + S_{\text{fin}}
    =
    \int\dd[3]{x}\left[
    (r_0 + \rs)^{3} + \frac{k^2 r_0}{2} - \frac{B^2}{r_0 +\rs}
    \right].
\end{equation}
The Euclidean version of the renormalized action can be regarded as the grand potential multiplied by $1/T$, thus the grand potential density is given by
\begin{equation}\label{eq:grand_potential}
	\Omega(T, \mu, k, B) =
	- (r_0 + \rs)^3 - \frac{k^2 r_0}{2}
	+
	\frac{B^2}{r_{0} + \rs}.
\end{equation}
This result agrees with those in the previous study \cite{Caldarelli:2016nni} for $B=0$.
We also obtain the energy density as
\begin{equation}
	\varepsilon(s, n, k, B) = \Omega + T s + \mu n
        = 2(r_0 + \rs)^3 - k^2 (r_0 + \rs) - \frac{k^2 \rs}{2}.
\end{equation}

The boundary value of the axio-dilaton can be read off as a vev of the dual operator, in the presence of the multi-trace deformation.
Using eq.~(\ref{eq:FG_expansions_axiodilaton}), we obtain the expression for the vev as
\begin{equation}\label{eq:O_sigma}
    \expval{O_{\sigma}}=\sigma_{(0)}
    =
    \frac{\rs}{2}\left(
    \frac{B^2-n^2}{B^2+n^2}
    - 2 i \frac{B n}{B^2 + n^2}
    \right).
\end{equation}
This expression depends on $n$ and $B$ via a ratio $n/B$.
Remarkably, the amplitude of the vev is simply given by
\begin{equation}
    \abs{\expval{O_{\sigma}}} = \frac{\abs{\rs}}{2}.
\end{equation}
Let us see the $T$-dependence of $\abs{\expval{O_{\sigma}}}$ by choosing $k$ as a scale.
The quantities we are interested can be written as
\begin{align}
    \frac{2\sqrt{2}\pi T}{k} = \frac{r_0(6r_0 + \rs)^2 - k^2}{2\sqrt{2}k\sqrt{r_0 (r_0 + \rs)}},\quad
    \frac{2\sqrt{2}\abs{\expval{O_{\sigma}}}}{k}
    = \frac{\sqrt{2}\abs{\rs}}{k}.
\end{align}
These two quantities are parameterized by two dimensionless parameters, such as $(k/r_0, \rs/r_0)$.
These two quantities are confined to a curve if we put one constraint on $n$ and $B$.
This fact implies the result does not depend on each $n$ and $B$ but a specific combination of them.
Actually, the result depends on $\sqrt{n^2 + B^2}$ only.
By scaling with $k$, it is written as
\begin{equation}
    \frac{\sqrt{n^2 + B^2}}{k^2} = \sqrt{\frac{3\rs(r_0 + \rs)(2(r_0+\rs)^2 - k^2)}{2 k^4}}.
\end{equation}
The explicit $B$-dependence is canceled out.
In figure \ref{fig:axion-dilaton_condensate}, we show $2\sqrt{2}\abs{\expval{O_{\sigma}}}/k$ as a function of the scaled temperature $2\sqrt{2}\pi T/k$ for various $\sqrt{n^2 + B^2}/k^2$.
As expected, the lower the temperature, the larger the order parameter would be obtained in general.
The result of the neutral limit given by eq.~(\ref{eq:O_sigma_neutral}) will be discussed in details later.

\subsection{Magnetization, susceptibility and thermodynamic relations}
\begin{figure}[htbp]
	\centering
	\includegraphics[width=14cm]{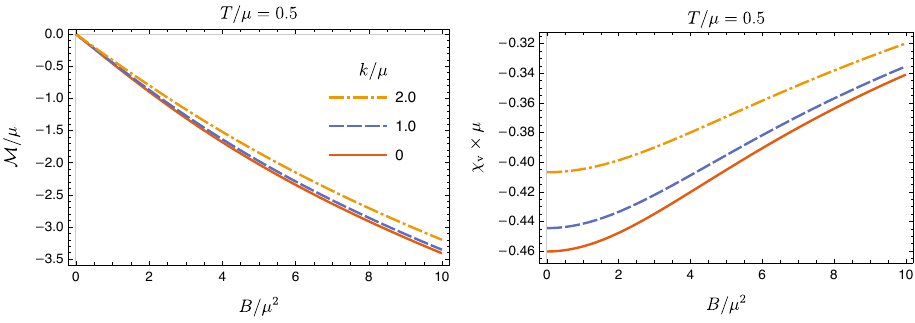}
	\caption{
        (left) $\mathcal{M}$ as a function of $B$,
        (right) $\chi_{\text{v}}$ as a function of $B$ for various $k$.
        We set $T/\mu = 0.5$.
	}
	\label{fig:M_chi_B}
\end{figure}

\begin{figure}[htbp]
	\centering
	\includegraphics[width=14cm]{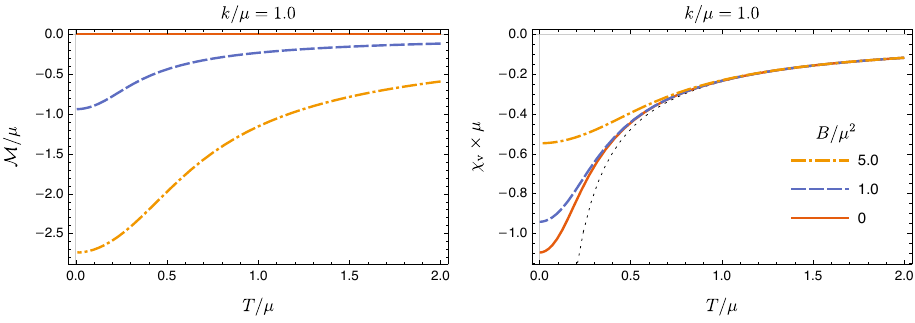}
	\caption{
        (left) $\mathcal{M}$ as a function of $T$, (right) $\chi_{\text{v}}$ as a function of $T$ for various $B$.
        We set $k/\mu=1.0$.
		In the right panel, the dotted curve denotes $\chi_{\text{v}} = - 3/(4\pi T)$.
	}
	\label{fig:M_chi_T}
\end{figure}

We are now ready to compute thermodynamic quantities from the thermodynamic potential.
The magnetization is obtained by differentiating the grand potential density (\ref{eq:grand_potential}) with respect to the magnetic field, which is given by%
\footnote{
In ref.~\cite{Seo:2015pug}, they discussed necessity of matching the dimension of the $(2+1)$d energy density to one in the $(3+1)$d theory, in order to compare the magnetization with actual measurement.
However, there is an ambiguity for choosing a dimensionful constant to match the dimensionality, so we do not follow their prescription here.
}
\begin{equation}
    \mathcal{M} :=
    - \left(
        \pdv{\Omega}{B}
    \right)_{T,\mu,k}
    =
    - \left(
        \pdv{\epsilon}{B}
    \right)_{s,n,k}
    =
    - \frac{B}{r_0 + \rs}.
\end{equation}
Note that the notation with subscripts means partial derivative under these values fixed.
One can perform such a partial differentiation by using the chain rule.
The magnetic susceptibility is obtained as
\begin{equation}\label{eq:magnetic_susceptibility}
    \chi_{\text{v}} := \frac{\mathcal{M}}{B}
    =
    - \frac{1}{r_{0} + \rs}.
\end{equation}
The high-temperature limit can be considered by taking large $r_{0}$, which gives
\begin{equation}
	\chi_{\text{v}} = - \frac{1}{r_{0}} + \order{r_{0}^{-2}}
	= - \frac{3}{4\pi T} + \order{T^{-2}}.
\end{equation}
Figure \ref{fig:M_chi_B} shows $\mathcal{M}$ and $\chi_{\text{v}}$ as functions of $B$ for fixed $T,k$, and Fig.~\ref{fig:M_chi_T} shows these as functions of $T$ for fixed $k,B$.
The analysis reveals that the magnetic susceptibility for both choices of parametrizations is always negative in this model. It's well known that negative susceptibility implies diamagnetism.
From the left panel of Fig.~\ref{fig:M_chi_B}, $\mathcal{M}$ monotonically decreases as $B$ increases.
In ref.~\cite{Kim:2015wba}, it was studied that the dyonic AdS solution without scalar hairs also shows similar behavior of diamagnetism.


Before closing this subsection, we check thermodynamic relations in this dyonic solution.
From the FG expansion of the metric (\ref{eq:FG_expansions_metric}), the nonzero components of the energy momentum tensor read
\begin{alignat}{5}
	\expval{T_{t}{}^{t}} &=&
	- &\frac{2}{3} \frac{B^2 + n^2}{\rs}
	+ \frac{k^2 \rs}{2}~
	&=&
	-&2(r_0 + \rs)^3 + k^2 (r_0 + \rs) + \frac{k^2 \rs}{2},\\
	\expval{T_{x}{}^{x}} &=&
	&\frac{1}{3} \frac{B^2 + n^2}{\rs}
	- \frac{k^2 \rs}{4}~
	&=&
	&(r_0 + \rs)^3 - \frac{k^2}{2}(r_0 +\rs) - \frac{k^2\rs}{4},
\end{alignat}
and $\expval{T_{y}{}^{y}} = \expval{T_{x}{}^{x}}$.%
\footnote{
	The result is not changed from their (5.52) for $\xi=3/2$ in ref.~\cite{Caldarelli:2016nni}.
}
The energy momentum tensor satisfies $\expval{T_{i}{}^{i}} =0$, and $\expval{T_{tt}} = - \expval{T_{t}{}^{t}} = \varepsilon$.
In ref.~\cite{Caldarelli:2016nni}, it was also mentioned that the equivalence between $\expval{T_{xx}}$ and the pressure $p$ no longer holds in the presence of the magnetic field and the axions.
We now define a total strength of the momentum dissipation by
\begin{equation}
	\Pi := \sqrt{\frac{1}{2}\sum_{I} (\partial_{i} \psi_I)^2}
	= k.
\end{equation}
Its thermodynamic conjugate is obtained as
\begin{equation}
	\varpi := - \left( \pdv{\Omega}{k}\right)_{T,\mu, B}\\
	=  k \left(
		2 r_{0} + \frac{3}{2} \rs
	\right).
\end{equation}
From the scaling analysis, as in \cite{Caldarelli:2016nni}, the equation of state is given by
\begin{equation}
	\varepsilon = 2 p - 2 \mathcal{M} B - \varpi \Pi,
\end{equation}
where $p$ denotes pressure, which is thermodynamic conjugate to the volume.
Using the equation of state, the pressure is obtained as
\begin{equation}
	p = (r_0+\rs)^3 + \frac{k^2 r_0}{2} - \frac{B^2}{r_0 + \rs} = - \Omega,
\end{equation}
which is an ordinary relation.
In terms of $\expval{T_{x}{}^{x}}$, we write the expression for $p$ as
\begin{equation}
	p = \expval{T_{x}{}^{x}} + \mathcal{M} B + \frac{1}{2} \varpi \Pi.
\end{equation}

\section{Axio-dilatonic neutral solution}\label{sec:neutral}
In this section, we consider a neutral limit of the dyonic solution while keeping a ratio between $n$ and $B$ to a specific value.
Rewriting $(n,B) = Q_{\rm EM}\times(\cos\xi,\sin\xi)$ in (\ref{eq:solution_1}), and taking $Q_{\rm EM} \to0$, we obtain a neutral solution parameterized by $\xi$:
\begin{equation}\label{eq:neutral_solution}
\begin{gathered}
	f(r) = h(r) \left(1 - \frac{k^2}{2(r+\rs)^2}\right),\quad
	h(r) = \sqrt{r(r+\rs)^3},\\
	e^{-\phi} = \frac{\sqrt{r(r+\rs)}}{r + \rs \sin^2\xi},\quad
	\chi = \frac{\rs \sin 2\xi}{-2 r - \rs + \rs \cos2\xi},
\end{gathered}
\end{equation}
and $A=0$.
For $\xi = 0$, it reduces to the known neutral dilatonic solution \cite{Zhou:2015qui} which is also studied in \cite{Ren:2021rhx,Jeong:2023ynk,Wang:2023rca}.
The horizon is located at $r=r_{0}$ where $r_{0}+\rs = k/\sqrt{2}$.
For general $\xi$, $\chi$ also has a non-trivial profile.
From the periodicity, we can restrict $\xi$ in $[0,\pi)$ without loss of generality.
The dimensionless Hawking temperature is given by
\begin{equation}
	T_k:=\frac{T}{k} = \frac{\sqrt{r_0(r_0 + \rs)}}{2\pi k}=\frac{1}{2\sqrt{2}\pi \sqrt{1+\xx}},
\end{equation}
where $\xx=r_b/r_0$. The grand potential density (\ref{eq:grand_potential}) becomes
\begin{equation}\label{eq:grand_potential_neutral}
	\Omega_k:=\frac{\Omega}{k^3} = - \frac{(r_0 + \rs)^2 (2 r_{0} + \rs)}{k^3}=-\frac{1}{2\sqrt{2}}(8\pi^2 T_k^2+1),
\end{equation}
which is independent of $\xi$.
Therefore, we cannot fix $\xi$ from thermodynamics.
Actually, $\xi$ is a phase of the complex variable $\expval{O_{\sigma}} = \phi_{(0)} + i \chi_{(0)}$.
From eq.~(\ref{eq:O_sigma}), we obtain
\begin{equation}
   \frac{ \expval{O_{\sigma}}}{k}
	= - \frac{\rs}{2k} e^{2 i \xi}
	= - \frac{1}{2\sqrt{2}}\left(
		1 - 8 \pi^2 T_k^2
	\right) e^{2 i\xi}.
\end{equation}
In addition, we can also understand $\xi$ as a parameter of the $U(1)$ symmetry (\ref{eq:U(1)_symmetry}).
If we perform the transformation (\ref{eq:U(1)_symmetry}) on the neutral solution (\ref{eq:neutral_solution}), we obtain the same solution with $\xi \to \xi + \theta$.
It is consistent with that the transformation corresponds to a rotation of $(n,B)$ in the charged case.

\begin{figure}[htbp]
	\centering
	\includegraphics[width=8cm]{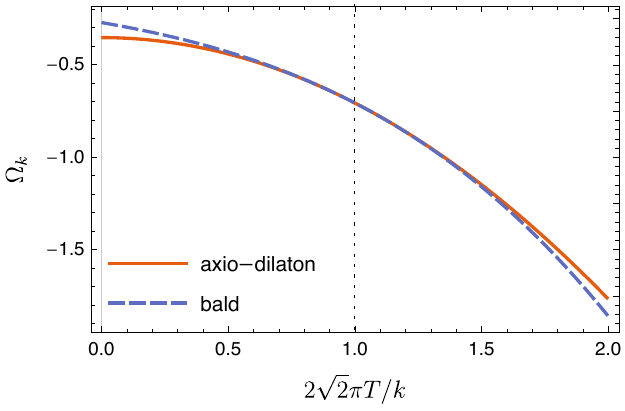}
	\caption{
		$\Omega_k$ as a function of $T_{k}$.
        The red-solid and the blue-dashed curves correspond to eqs.~(\ref{eq:grand_potential_neutral}) and (\ref{eq:grand_potential_neutral_bald}), respectively.
        These curves crosses at $2\sqrt{2}\pi T/k = 1$.
	}
	\label{fig:energy_difference}
\end{figure}

\begin{figure}[htbp]
	\centering
	\includegraphics[width=10cm]{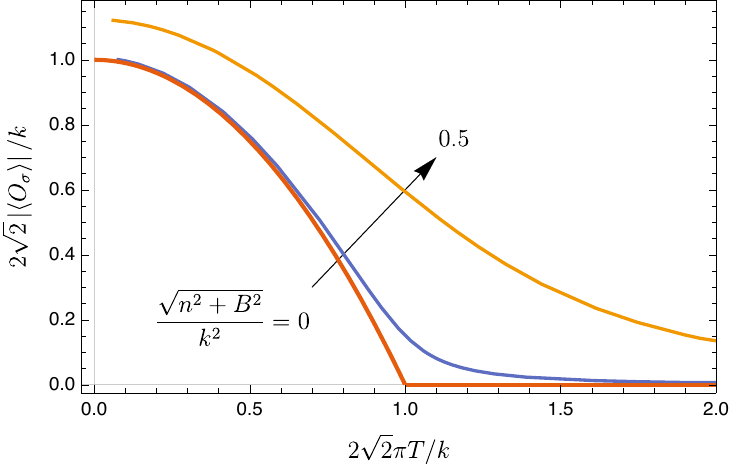}
	\caption{
		$2\sqrt{2}\abs{\expval{O_{\sigma}}}/k$ as a function of $2\sqrt{2}\pi T / k$ for various $\sqrt{n^2+B^2}/k^2$.
        The curves correspond to $\sqrt{n^2+B^2}/k^2=0, 0.1, 0.5$ from bottom to top.
        In particular, the curve for the neutral case is given by eq.~(\ref{eq:O_sigma_neutral}).
	}
	\label{fig:axion-dilaton_condensate}
\end{figure}

On the other hand, the system admits another neutral black brane solution.
It is given by \cite{Andrade:2013gsa}
\begin{equation}
    f(r)=h(r)\left[
        1-\frac{k^2}{2r^2}-\frac{r_0^3}{r^3}\left(
            1-\frac{k^2}{2r_0^2}
        \right)
    \right],\quad
    h(r)=r^2,\quad e^{-\phi}=1,\quad \chi =0,
\end{equation}
and $A=0$.
This is a so-called bald solution without axio-dilaton hairs.
The Hawking temperature is given by $T=\frac{1}{8\pi}\frac{6 r_0^2 - k^2}{r_0}$.
The grand potential of this solution is obtained as
\begin{align}\label{eq:grand_potential_neutral_bald}
\begin{split}
    \Omega_k &= - \frac{r_{0}^3}{k^3}\left(1+\frac{k^2}{2r_{0}^2}\right)\\
    &=-\frac{1}{54}\left(\sqrt{16 \pi^2 T_k^2+6}+4 \pi T_k\right)^3\\
    &+\frac{1}{9} \pi T_k\left(\sqrt{16 \pi^2 T_k^2+6}
    +4 \pi T_k\right)^2
    \end{split}
\end{align}
Figure \ref{fig:energy_difference} shows $\Omega_k$ as a function of $T_k$ for each solution.
Comparing eqs.~(\ref{eq:grand_potential_neutral}) and (\ref{eq:grand_potential_neutral_bald}), one can find that the bald solution is thermodynamically favored at high temperatures when $T>k/(2\sqrt{2}\pi)$, as the same as in \cite{Jeong:2023ynk,Wang:2023rca}.
In this solution, $\sigma_{(0)}$ vanishes.
As a result, the vev of the complex operator $O_{\sigma}$ behaves as
\begin{equation}\label{eq:O_sigma_neutral}
   \frac{ \expval{O_{\sigma}}}{k} =
    - \frac{1}{2\sqrt{2}}
    \left(
	       1 - 8 \pi^2 T_k^2
	\right) e^{2 i\xi}
    \times
    \Theta\left(1-2\sqrt{2}\pi T_k\right),
\end{equation}
where $\Theta(x)$ denotes the step function.
Figure \ref{fig:axion-dilaton_condensate} shows the behavior of $\abs{\expval{O_{\sigma}}}$ as a function of $2\sqrt{2}\pi T/k$ for various values of $\sqrt{n^2+B^2}/k^2$.
The neutral result for $\sqrt{n^2 + B^2}/k^2 = 0$ is expressed by eq.~(\ref{eq:O_sigma_neutral}).
Remarkably, $\expval{O_{\sigma}}$ exhibits the second order phase transition at $T=k/(2\sqrt{2}\pi)$, and nonzero $\expval{O_{\sigma}}$ at low temperature implies the spontaneous breaking of the global $U(1)$ symmetry.
From eq.~(\ref{eq:O_sigma_neutral}), we can write
\begin{equation}
    \expval{O_{\sigma}} \sim 2 \pi e^{2i\xi} (T-\hat{T}_{c}),
\end{equation}
near $T=\hat{T}_{c}=k/(2\sqrt{2}\pi)$, so the critical exponent can be read off as $1$ obviously.
It differs from a typical value of $1/2$ in most of holographic studies on the second order PT, and also in the mean field theory.
For finite charge cases, the phase transition behavior is smoothed.

The phase $\xi$ is also related to the DC conductivities and the Hall angle.
From eq.~(\ref{eq:DC_conductivities}) for the axio-dilatonic solution, each component of the DC conductivity are written as
\begin{align}\label{eq:DC_cond_neutral}
    \begin{split}
        \sigma_{xx} &= e^{-\phi(r_0)}
	= \frac{2\sqrt{r_0 (r_0+\rs)}}{2 r_0 + \rs -\rs \cos2\xi}=\frac{4 \sqrt{2} \pi T_k}{\left(8 \pi^2 T_k^2-1\right) \cos 2 \xi+8 \pi^2 T_k^2+1},\\
	\sigma_{xy} &= - \chi(r_0)
	= \frac{\rs \sin2\xi}{2 r_0 + \rs - \rs \cos2\xi}=\frac{\left(1-8 \pi^2 T_k^2\right) \sin 2 \xi}{\left(8 \pi^2 T_k^2-1\right) \cos 2 \xi+8 \pi^2 T_k^2+1},
    \end{split}
\end{align}
and $\sigma_{yy}=\sigma_{xx}$, $\sigma_{yx}=-\sigma_{xy}$.
From Eq.~(\ref{eq:DC_cond_neutral}), it is straightforward to see that the non-zero Hall angle directly results from the non-vanishing axion in the bulk, and the $\chi F\tilde{F}$ term in the action.
%
For the bald solution, $\sigma_{xy}=0$ since $\chi $ vanishes. Interestingly, $\sigma_{xy}=\cot\xi$ is non-vanishing for generic $\xi$ in the $T_k\to 0$ limit and $\sigma_{xy}\to0$ as $T_k\to T_k^c=1/(2\sqrt{2}\pi)$, which is consistent with the phase transition picture.
From Eq.~(\ref{eq:Hall_angle}), the Hall angle becomes
\begin{equation}
\begin{aligned}
	\tan\theta_{H}
    =& \frac{\rs}{2\sqrt{r_0(r_0 +\rs)}} \sin2\xi
    \times
    \Theta\left(1-2\sqrt{2}\pi T_k\right)\\
    =&
    \frac{1}{4\sqrt{2}\pi T_k}\left(
        1- 8\pi^2 T_k^2
    \right)\sin2\xi
    \times
    \Theta\left(1-2\sqrt{2}\pi T_k\right),
\end{aligned}
\end{equation}
where the step function is multiplied to reflect the phase transition.
The non-zero Hall conductivity in the absence of the magnetic field is similar to the anomalous Hall effect (AHE).
The AHE is triggered by the spontaneous magnetization in  ferromagnetic or antiferromagnetic materials, but our system has no spontaneous magnetization.
In our case, therefore, there might be another origin to trigger the Hall effect without the magnetic field.

\begin{figure}[htbp]
	\centering
	\includegraphics[width=10cm]{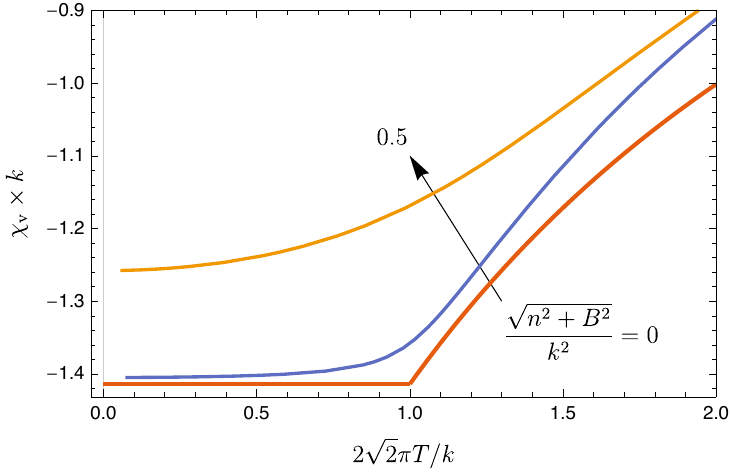}
	\caption{
		$\chi_{\mathrm{v}}\times k$ as a function of $T/k$ for various $\sqrt{n^2+B^2}/k^2$.
        The curves correspond to $\sqrt{n^2+B^2}/k^2=0,0.1,0.5$ from bottom to top.
        The neutral result is given by eq.~(\ref{eq:chiv_neutral}).
	}
	\label{fig:chi_k}
\end{figure}

The magnetic susceptibility remains finite in the neutral limit because it is a linear response coefficient.
From eq.~(\ref{eq:magnetic_susceptibility}), we obtain $\chi_{\mathrm{v}}=-\sqrt{2}/k$.
From the results studied in ref.~\cite{Kim:2015wba} or \cite{Seo:2015pug}, the magnetic susceptibility of the bald solution is given by $\chi_{\mathrm{v}}=-1/r_{0}$.
Taking into account the phase transition, we obtain
\begin{equation}\label{eq:chiv_neutral}
  \chi_{\mathrm{v}} \times k =
   -\sqrt{2}\times
   \begin{cases}
      1 & T_k \leq \frac{1}{2\sqrt{2}\pi}\\
      \frac{3}{2\sqrt{2}\pi T_k + \sqrt{3 + 8\pi^2 T_k^2}} &
      \frac{1}{2\sqrt{2}\pi} < T_k
   \end{cases}.
\end{equation}
The temperature behavior of $\chi_{\mathrm{v}}$ with setting $k$ as a scale is shown in figure \ref{fig:chi_k}.
The constant susceptibility in the low temperature region agrees with a typical behavior of diamagnetism, e.g., see chapter 11 in \cite{kittel2018introduction}.
Similar to Fig.~\ref{fig:axion-dilaton_condensate}, the curve is smoothed when we consider finite $\sqrt{n^2 + B^2}/k^2$.

\subsection{Analytic AC conductivity in the neutral solution}
In this subsection, we compute the AC conductivity for the neutral solution.
We find an analytic form of the AC conductivity in terms of Heun functions.
Our result is a natural extension of the result obtained in \cite{Ren:2021rhx} to the S-completed model.
\rev{
Note that the current operator dual to $A_{\mu}$ do not represent the conserved current associated with the global $U(1)$ symmetry (\ref{eq:U(1)_symmetry}), which can be spontaneously broken in the neutral limit.
Here, we consider the action (\ref{eq:S-completed}) for the coupling between the vector fields and the axio-dilaton.
}

Let us now consider the vector perturbation $A = A_{x}(t, r) \dd{x} + A_{y}(t,r)\dd{y}$ to compute the AC conductivity.
Note that due to the Chern-Simons coupling, $A_{x}$ and $A_{y}$ are coupled.
The linearized equations of motion are given by
\begin{equation}
\begin{aligned}
	\omega \frac{e^{\phi}\chi'}{f} A_{y}(r)
	+ \frac{\omega^2}{f^2} A_{x}(r)
	+ \left(\frac{f'}{f} - \phi'(r)\right) A_{x}'(r) + A_{x}''(r) &= 0,\\
	- \omega \frac{e^{\phi}\chi'}{f} A_{x}(r)
	+ \frac{\omega^2}{f^2} A_{y}(r)
	+ \left(\frac{f'}{f} - \phi'(r)\right) A_{y}'(r) + A_{y}''(r) &= 0,
\end{aligned}
\end{equation}
in the Fourier space.
Rewriting $A_{\pm} = A_{x} \pm i A_{y}$, we can decouple the equation as
\begin{equation}
	\left(
		\frac{\omega^2}{f^2} \pm \omega \frac{e^{\phi}\chi'}{f}
	\right) A_{\pm}(r)
	+ \left(
		\frac{f'}{f} - \phi'
	\right) A_{\pm}'(r) + A_{\pm}''(r) =0.
\end{equation}
With the background solution (\ref{eq:neutral_solution}), these ODEs have four regular singular points at $r = 0, r_0, -(r_0 + 2 \rs),$ and  $- \rs \sin^2\xi$.
The solution can be written in terms of the Heun polynomial.
The ingoing-wave solution is written as
\begin{equation}
	A_{\pm}(r) = \mathcal{C}_{\pm} u^{-i\ws/2}
	\heun\left(
		\tau, q_{\pm}; \alpha, \beta, \gamma , \delta; u
	\right),
	\quad
	u = 1 - \frac{2 (1+ \xx) r}{r + r_{0} (1 + 2 \xx)},
\end{equation}
where $\mathcal{C}_{\pm}$ are normalization constants, $\xx=\rs/r_0$, $\ws = \omega/(2\pi T)$, and the parameters are given by
\begin{equation}
\begin{gathered}
	\tau = -\frac{
        (1+2\xx)(-2 + \xx(-1 + \cos2\xi))
    }{2 + \xx (3 + \cos2\xi)},\\
	q_{\pm} =
	\frac{
		2 \xx \sin\xi \left(
			\pm \sqrt{\xx+1}\cos\xi - i (\xx+1) \sin\xi
		\right)
	}{\xx \cos2\xi +3 \xx+2} \ws
	- \frac{\xx (1 + \xx \sin^2\xi)}{2 + 3 \xx + \xx \cos2\xi} \ws^2,\\
	\alpha = \frac{\ws}{2}\left(
		- i + \frac{1}{\sqrt{-1-2 \xx}}
	\right),\quad
	\beta = \frac{\ws}{2}\left(
		- i - \frac{1}{\sqrt{-1-2 \xx}}
	\right),\quad
	\gamma = 1 - i \ws,\quad
	\delta = 1.
\end{gathered}
\end{equation}
The double sign appears only in the accessory parameter $q_{\pm}$.
Note that Heun polynomial satisfies
\begin{equation}
	\left[
		\frac{\alpha \beta u - q}{u(u-1)(u - \tau)}
		+\left(
			\frac{\gamma}{u} + \frac{\delta}{u-1} + \frac{\epsilon}{u - \tau}
		\right)\frac{\partial}{\partial u}
		+
		\frac{\partial^2}{\partial u^2}
	\right] \heun (u) =0,
\end{equation}
with regular condition at $u=0$.
The location of horizon and boundary, $r=r_0, \infty$, are mapped to $u=0, -  1 - 2\xx$, respectively.
In the limit $\xi =0$, we obtain $\tau=1$ and $\epsilon = -1$.
In this case, the number of the regular singularities reduces to $3$, then the solution reduces to hypergeometric function.

Another form of the Heun's equation is given by, (e.g., see ref.~\cite{Ge:2024jdx},)
\begin{equation}\label{eq:Heun_eq_standard_2}
\begin{aligned}
    \psi''(u) + \Bigg[&
        \frac{\frac{1}{4}-a_{0}^2}{u^2}
        + \frac{\frac{1}{4}-a_{1}^2}{(u-1)^2}
        + \frac{\frac{1}{4}-a_{\tau}^2}{(u-\tau)^2} +\\
        &- \frac{\frac{1}{2} - a_{0}^2 -a_{1}^2 - a_{\tau}^2 + a_{\infty}^2 + \bar{u}}{u(u-1)}
        + \frac{\bar{u}}{u(u-\tau)}
    \Bigg]\psi(u)
    =0.
\end{aligned}
\end{equation}
The parameters are obtained as
\begin{gather}
    a_{0} = \frac{i\ws}{2},\quad
    a_{1} = 0,\quad
    a_{\tau} = 1,\quad
    a_{\infty} = - \frac{i \ws}{2\sqrt{1 + 2 \xx}},\\
    \bar{u} =
    \frac{\xx(3+4\xx) + (1 +2\xx)\csc^2\xi}{4\xx(1+\xx)}
    \mp \frac{|\sin2\xi|\csc^2\xi}{4\sqrt{1+\xx}}\ws,
\end{gather}
The function $\psi$ is related to $\heun$ by
\begin{equation}
    \heun(u) = (1-u)^{-\frac{\delta}{2}} u^{-\frac{\gamma}{2}} (\tau-u)^{-\frac{\epsilon}{2}} \psi(u).
\end{equation}

\begin{figure}[htbp]
	\centering
	\includegraphics[width=14cm]{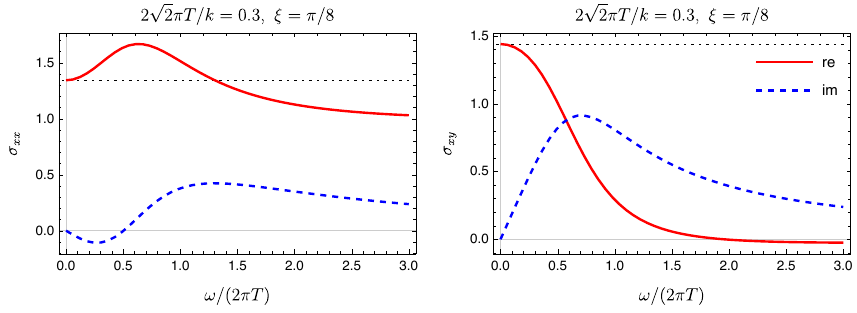}
	\caption{
        AC conductivities for $2\sqrt{2}\pi T/k =0.3$, $\xi = \pi/8$.
        The horizontal dotted lines denotes the analytic DC values.
	}
	\label{fig:AC_cond_neutral}
\end{figure}

We can formally write the analytic expression of the AC conductivities in terms of the Heun function, as follows.
Each component of the gauge fields can be obtained by $A_{x} = (A_{+} + A_{-})/2$ and $A_{y} = (A_{+} - A_{-})/(2i)$.
Note that the normalization constants $\mathcal{C}_{\pm}$ are not fixed yet.
Their ratio $c_{\pm}:=\mathcal{C}_{-}/\mathcal{C}_{+}$ can be fixed by imposing the Dirichlet condition by choosing $\lim_{r\to\infty}A_{x}(r) = 0$ or $\lim_{r\to\infty}A_{y}(r) = 0$.
We write
\begin{align}
	A_{x} =& \frac{\mathcal{C}_{+}}{2} u^{-i\ws/2}\left[
		\heun(q_{+};u)
		+ \frac{\mathcal{C}_{-}}{\mathcal{C}_{+}}
		\heun(q_{-}; u)
	\right],\\
	A_{y} =&
	\frac{\mathcal{C}_{+}}{2i} u^{-i\ws/2}\left[
		\heun(q_{+}; u)
		- \frac{\mathcal{C}_{-}}{\mathcal{C}_{+}}
		\heun(q_{-}; u)
	\right].
\end{align}
where
\begin{equation}
    c_{\pm}=
	\frac{\mathcal{C}_{-}}{\mathcal{C}_{+}}
	=
	\pm
	\frac{
		\heun(q_{+}; u_{\infty})
	}{
		\heun(q_{-}; u_{\infty})
	},\quad
	u_{\infty} := - 1 - 2 \rs.
\end{equation}
The positive sign corresponds to imposing $\lim_{r\to\infty} A_{y} = 0$, while the negative sign corresponds to $\lim_{r\to\infty} A_{x} = 0$.
Using the Kubo formula $\sigma(\omega)=G^{\rm R}(\omega)/i\omega$, the $x$--$x$ component of the conductivity is obtained as
\begin{equation}
\begin{aligned}
	\sigma_{xx}(\omega)
	=&
	\frac{1}{i\omega}\left.\lim_{r\to\infty} \frac{\partial_{r} A_{x}(r)}{A_{x}(r)}\right|_{E_{y}=0}
	=
	\sqrt{1+\xx}
	+ \frac{1}{i\omega}\lim_{r\to\infty}\frac{
		\partial_{r}[\heun(q_{+}; u) + c_{+} \heun(q_{-}; u)]
	}{
		\heun(q_{+}; u) + c_{+} \heun(q_{-}; u)
	}.
\end{aligned}
\end{equation}
We have used $\ws = \omega/(r_0 \sqrt{1+\xx})$ here.
The $x$--$y$ component is obtained as
\begin{equation}
\begin{aligned}
	\sigma_{xy}(\omega) =&
	\frac{1}{i\omega} \left.
	\lim_{r\to\infty} \frac{\partial_{r} A_{x}(r)}{A_{y}(r)}
	\right|_{E_{x} = 0}
	=
	\frac{1}{\omega}
	\lim_{r\to\infty}
	\frac{
		\partial_{r}\left[
			\heun(q_{+}; u) + c_{-} \heun(q_{+}; u)
		\right]
	}{
		\heun(q_{+}; u) - c_{-} \heun(q_{+}; u)
	}.
\end{aligned}
\end{equation}
One can also write $\sigma_{yy}$ and $\sigma_{yx}$ as in similar forms.
Figure \ref{fig:AC_cond_neutral} shows the numerical results of the AC conductivities $\sigma_{xx}$ and $\sigma_{xy}$ for $\xx = 10$, $\xi = \pi/8$.
Note that there is no Dirac delta in the AC conductivity because the background is neutral, and the condensate $\expval{O_{\sigma}}$ is not charged under the electromagnetic $U(1)$ symmetry.
The DC values agree with Eq.~(\ref{eq:DC_cond_neutral}).
The AC Hall conductivity $\sigma_{xy}(\omega)$ vanishes when $\chi$ vanishes, i.e., $\sin2\xi=0$.

\begin{figure}[htbp]
	\centering
	\includegraphics[width=8cm]{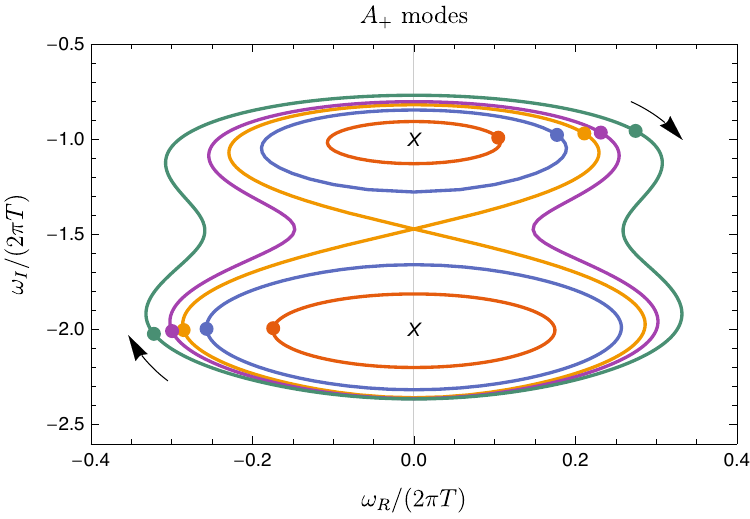}
	\caption{
		Quasinormal frequencies of the lowest two modes for $A_{+}$.
		The curves denote periodic trajectory parameterized by $0<\xi<\pi$, for various $\xx$.
        The dots on the curves denote mode frequencies at $\xi = \pi/4$. The curves correspond to $\xx = 0.25, 0.5, 0.6483, 0.75, 1.0$ from inside to outside.
        The corresponding temperatures are $2\sqrt{2}\pi T_k = 0.89443, 0.81650, 0.77890, 0.75593, 0.70711$, respectively.
        The two modes share the periodic trajectory when $\xx \geq 0.6483$ corresponding to $2\sqrt{2}\pi T_k \leq 0.77890$.
		The cross markers denote $\omega = - 2 i \pi n T$.
	}
	\label{fig:QNMs_trajectory}
\end{figure}

Figure \ref{fig:QNMs_trajectory} shows the location of the quasinormal modes (QNMs) for the $A_{+}$ fluctuation in the complex $\omega$-plane.
The result for $A_{-}$ modes is the same by flipping $\omega_{R} \to - \omega_{R}$.
These complex-valued frequencies correspond to the poles of the AC conductivity tensor.
The curves are trajectories parametrized by $\xi$ for various $\xx$.
We show the first and the second spectra at $\xi=\pi/4$ as dots.
When $\xi=0$, all spectra are located on the imaginary axis.
As we increase $\xi$, the spectra move clockwise around $\omega = - 2i \pi n T$ corresponding to the zeros of the Green's function.
For large $\xx$, two modes have a common trajectory involving two zeroes.
The trajectory has pits at $\xi=\pm \pi/2$ for relatively small $\xx$ around one.
Furthermore, the trajectories of two modes separate below a specific critical value of $\xx$ around $\xx = 0.6483$.
The corresponding scaled-temperature is given by $2\sqrt{2}\pi T_{k} = 0.77890$.
At this critical value, the trajectory has a crossing point at $\omega/2\pi T = - 1.4707 i$, and the two modes collide when $\xi=\pi/2$.
Below the critical value of $\xx$, each mode moves around each zero point by varying $\xi$.
As $\xx$ further decreases, each trajectory shrinks.
It is expected that all pole and zero cancel with each other when $\xx$ reaches zero or when $2\sqrt{2}\pi T_{k}$ reaches one.
The behavior of the first pole reflects the peak at nonzero $\omega$ in $\sigma_{xx}(\omega)$.
The behavior of the QNMs is similar to those studied in ref.~\cite{Kim:2015wba} with finite electric and magnetic charges.
However, we cannot interpret these as cyclotron bound states since we are dealing with the neutral limit.
The presence of the QNMs having nonzero $\omega_{R}$ for $\xi\neq0$ obviously reflects the nonzero Hall conductivity.

\section{Conclusions and discussions}\label{sec:discussion}
In this paper, we provide the expression of the thermodynamic potential of the axio-dilaton dyonic black brane in the S-completed Gubser-Rocha model with the translational symmetry breaking axions.
We also provide the analytic expressions of the magnetization and the magnetic susceptibilities by using the thermodynamic potential.
As a result, we find the system always exhibits diamagnetism.
In order to interpret the black brane solution as a vacuum state, we consider the triple trace marginal deformation by adding appropriate boundary action.
In this treatment, the leading term of the axio-dilaton in the vicinity of the AdS boundary can be identified with the corresponding operator's vev with scaling dimension $\Delta_{-}=1$.
The vev of the dual operator to the axio-dilaton can be written in the complex-valued expression, and its amplitude is given by the characteristic scale $\rs$ of the dyonic solution.

In the latter part, we consider a neutral limit while keeping the ratio between the electric and the magnetic charge of the dyonic black brane fixed.
This limit gives the non-trivial neutral solution with one parameter $\xi$ corresponding to the ratio.
Interestingly, $\xi$ is directly related to the phase of the complex operator which is a dual to the axio-dilaton field, and it is also related to the unconventional Hall effect.
\rev{
Moreover, the thermodynamic energy of the neutral solution does not depend on the choice of $\xi$, and $\xi$ is a parameter of the global $U(1)$ symmetry (\ref{eq:U(1)_symmetry}) of the bulk action.
}%
As a result, we interpret the condensation of the complex operator as spontaneous breaking of a global $U(1)$ symmetry in the boundary theory.
It can be considered as a second order phase transition, and the operator's vev can be written in an analytic form.
For this picture, the presence of the axio-dilaton and the marginal deformation are essential.
In the absence of the axion field, the symmetry breaking becomes $Z_2$ because the dual operator to the dilaton is restricted to the real axis.
By considering the gauge field fluctuations, we also obtain the analytic expression of the AC conductivities in terms of the Heun's polynomials in the neutral limit. 

We remark some points on the global $U(1)$ symmetry breaking we observed in the neutral limit of our model.
In our model, the axio-dilaton is not charged in the bulk.
This fact corresponds to the complex operator has also no charge under any gauge transformation.
\rev{
In fact, the global symmetry can be found in the bulk action as a $SO(2)\simeq U(1)$ subgroup of $SL(2;\mathbb{R})$.
The AdS/CFT dictionary usually states that a local symmetry in the bulk corresponds to a global symmetry in the boundary theory, and the gauge fields are dual fields to the Noether's currents.
In our case, however, the bulk symmetry is global, and there is no gauge field.
It should be clarified what actually happens in the boundary theory in this case.
We also remark that the $SO(2)$ subgroup is modded out in the context of higher-dimensional supergravity \cite{Bergshoeff:2002nv} and at the quantum level this symmetry must be absent due to swampland no global symmetry conjecture. Therefore this symmetry should be understood as an approximate symmetry and it is expected to be broken to $\mathbb{Z}_4$ non-perturbatively when quantum corrections are included.
}

Each component of the conductivity tensor depends on the value of the phase $\xi$.
It leads to the unconventional Hall effect that remains finite in the absence of the magnetic field.
As we have mentioned, the non-vanishing Hall effect is typically observed in antiferromagnetism or ferromagnetism, and it is called AHE.
However, our system exhibits diamagnetism with no spontaneous magnetization, so it  differs from the AHE.
We note that our unconventional Hall effect can be observed only when we take both electric and magnetic charges as zero.
In our case, the phase of the complex operator $\xi$, which is related to $n/B$ originally, may play the role of an effective magnetic field.
Although the setup is very different from ours, a similar unconventional Hall effect had been observed in a non-Abelian holographic SC model \cite{Roberts:2008ns}.
They observed a mode with a Dirac delta at a finite frequency in the AC conductivity spectrum.
Our results of the AC conductivities also indicate a similar mode at a finite frequency, which also has a finite decay width since our system is not a superconductor.
The understanding of these modes would be important to understand the dual description of this system.

We can naturally expect the emergence of a Goldstone mode with the $U(1)$ symmetry breaking in the neutral limit.
On the bulk side, it should be described by the quasinormal mode of the axio-dilaton fluctuation around the black brane solution.
\rev{%
Again, we remark that our case is different from that of the spontaneous breaking of the bulk local symmetry, such as in usual holographic superfluid models.
However, there are some examples that the spontaneous breaking of a bulk global symmetry also results in Goldstone modes in holography.
For instance, the emergence of the pion-like excitation, which is a Goldstone mode of the chiral symmetry breaking, was studied in the non-supersymmetric D3-D7 model in \cite{Babington:2003vm}.
In \cite{Amado:2013xya}, they studied the emergence of Goldstone modes in the ungauged model, which exhibits spontaneous breaking of a local $U(1)$ symmetry and a global $SU(2)$ symmetry.
It would also be interesting to analyze which type of Goldstone modes we have for our model and reveal its physical meaning by analyzing the dispersion relation in detail.
}%


\revb{
The mechanism of spontaneous symmetry breaking in this study is similar to those studied in \cite{Ren:2022qkr} and in Sec.~2.3 of \cite{Faulkner:2010gj} for vanishing density.
In these cases, spontaneous symmetry breaking is caused by the double-trace operator deformation and the presence of the nonlinear scalar potential.
One of the main differences is that the scalar is charged under a bulk local $U(1)$ symmetry in their models.
Another difference is that we consider the marginal deformation rather than the relevant deformation, and the linear axions that give the scale $k$.
In the presence of the linear axions, we can obtain the analytic black brane solution with finite temperature.
In addition, our model contains the axio-dilaton field that transforms under modular $SO(2)$.
It leads to a dual complex operator, which can be interpreted as the order parameter of the global $U(1)$ symmetry breaking. 
}

In the neutral limit, we obtain the analytic AC conductivities.
Recently, the connection between QNM and quantum Seiberg-Witten curve \cite{Aminov:2020yma,Bonelli:2021uvf,Bianchi:2021mft} has been studied.
By mapping Heun's equation to a problem of the quantum SW curve, one can obtain a connection formula and write the holographic Green's function or the AC conductivity in analytic series if the expansion is performed around the singular points of the Heun function.
In our case, however, we cannot apply this method to our study because the AdS boundary is not a regular singular point of the Heun's equation for computing the AC conductivity.
This problem happens for a vector perturbation in asymptotic AdS$_4$ spacetime in general. It would be interesting to see whether the techniques developed in \cite{Aminov:2023jve} can be applied to the analytic study of the QNM of our model.

The solution studied in this paper exhibits finite Hall conductivity even in the absence of the magnetic field.
Such a property originate from the $\chi F\wedge F$ term in the action of the model, similar to the models studied in \cite{Goldstein:2010aw,Seo:2015pug}.
This term may lead to a Chern-Simons (CS) term like $\chi_{(0)} A \wedge \dd{A}$ in the $(2+1)$d boundary theory.
On the other hand, such a term in holographic models can induce a striped instability \cite{Nakamura:2009tf,Bergman:2011rf}.
A similar striped instability would be observed in our model probably.
In our model, the axion is coupled in the CS term additionally.
Investigating the influence of the non-trivial axio-dilaton field to the instability would be worth studying.

It would be interesting to consider dynamical Maxwell fields in the boundary field theory \cite{Montull:2009fe,Domenech:2010nf,Maeda:2010br}.
Similar to \cite{Jeong:2023las, Baggioli:2023oxa}, we can introduce the dynamical Maxwell fields by adding the boundary kinetic term such as
\begin{equation}
    S_{\rm EM} := \int\dd[3]{x}\left(
        - \frac{1}{4\lambda} F_{ij}F^{ij} + A_{i} J^{i}_{\rm ext}
    \right),
\end{equation}
where $\lambda$ is a coupling constant of dimension $1$, and $J^{i}_{\rm ext}$ is the external source to the Maxwell fields.
With this finite boundary action, the Maxwell equation emerges in the boundary theory as a boundary condition for $A_{\mu}$.
With $J^{i}_{\rm ext}=0$, the dyonic solution may admit that boundary condition.
The additional boundary action, therefore, just gives a contribution of $B^2/(2\lambda)$ to the grand potential density.
It corresponds to the energy density of the magnetic field itself in the real measurement.
The presence of the dynamical Maxwell fields in the boundary theory would play more nontrivial roles when the system has a finite condensation charged under the gauge symmetry, i.e., in the so-called bona-fide holographic superconductor.
In such cases, we expect that the system exhibits a state with magnetic field vortices, which may be interpreted as the Abrikosov vortex.
Many related works on the superconudcting phase under the  magnetic fields had been conducted in the Hartnoll-Herzog-Horowitz (HHH) model as in~\cite{Albash:2009iq,Tallarita:2019amp,Xia:2021jzh}.
The inclusion of the dynamical Maxwell field would also change the spectrum of the system as those studied in ref.~\cite{Ahn:2022azl}.
It would be interesting to study such phenomena in the holographic superconductor based on the S-type Gubser-Rocha model studied here, but we leave it for future work.%
\footnote{
We present the phase diagrams of the holographic superfluid based on our model, in Appendix \ref{appendix:charged_sector}.
}

\acknowledgments
We would like to thank Xian-Hui Ge, Keun-Young Kim, Kilar Zhang and Rui-Dong Zhu for helpful discussions.

\appendix
\section{Charged scalar instability under the external magnetic field}\label{appendix:charged_sector}
We consider a charged scalar sector to investigate the phase diagram of its instability under the external magnetic field, here.
The Lagrangian density is given by
\begin{equation}
	\frac{\mathcal{L}_{3}}{\sqrt{-g}} = - |D_{\mu} \Psi|^2 - M^2 |\Psi|^2,\quad
	D_{\mu} = \nabla_{\mu} - i q A_{\mu},
\end{equation}
where $M$ is a mass of the scalar field, and $q$ is a charge.
We will set $M^2=-2$ and $q=6$.
The total action becomes $S=\frac{1}{2\kappa}\int\dd[4]{x}\sqrt{-g}(\mathcal{L}_{1}+\mathcal{L}_{2}+\mathcal{L}_{3})$.
The boundary theory's setup is considered as an s-wave superfluid immersed in an external magnetic field.
For our purpose, we only consider the linear order fluctuation of the charged scalar field.
The presence of the external magnetic field induces instability of an inhomogeneous phase of the charged scalar condensation.

The equation of motion for the charged scalar field can be written as
\begin{equation}
	\frac{1}{\sqrt{-g}}\partial_{\mu} \left[ \sqrt{-g} \left(
		 \partial^{\mu}\Psi - i q A^{\mu}\Psi\right)
	\right]
	- i q A_{\mu} \partial^{\mu}\Psi
	- q^2 A_{\mu} A^{\mu}\Psi - M^2\Psi=0.
\end{equation}
If the scalar field's perturbation is small, it is decoupled from the other perturbations up to the first order.
In a polar coordinate introduced by $(x,y)=(R\cos\theta,R\sin\theta)$, the metric becomes
\begin{equation}
	\dd{s}^2 = -f(r) \dd{t}^2 + \frac{\dd{r}^2}{f(r)}
	+ h(r)(\dd{R}^2 + R^2 \dd{\theta}^2),
\end{equation}
and the gauge potential in the symmetric gauge becomes
\begin{equation}
	A = A_0 \dd{t} + \frac{B}{2} R^2 \dd{\theta}.
\end{equation}
In terms of the metric components, the equation is written as
\begin{equation}
	\partial_{r}^2\Psi
	+ \left(
		\frac{f'}{f} + \frac{h'}{h}
	\right) \partial_{r} \Psi
	+ \frac{f^{-1} q^2 A_0 - M^2}{f}\Psi
	+ \frac{1}{f h} \left[
		\partial_{R}^2 + \frac{1}{R} \partial_{R}
		- \frac{q^2}{4} B^2 R^2
	\right] \Psi = 0.
\end{equation}
A similar equation appears in the studies of the HHH model under the magnetic fields \cite{Albash:2008eh, Hartnoll:2008kx, Ge:2010aa}.
Following \cite{Hartnoll:2008kx}, we assume a separable ansatz for $\Psi$ as $\Psi(r, R) = X(R) F(r)$.
The equation becomes
\begin{equation}
	\partial_{r}^2 F
	+ \left(
		\frac{f'}{f} + \frac{h'}{h}
	\right) \partial_{r} F
	+ \frac{f^{-1} q^2 A_0 - M^2}{f}F
	+ \frac{1}{f h}
	\frac{F}{X}
	\left[
		\partial_{R}^2 + \frac{1}{R} \partial_{R}
		- \frac{q^2}{4} B^2 R^2
	\right] X = 0.
\end{equation}
We can separate the equation as
\begin{gather}
	\left[
		\partial_{R}^2 + \frac{1}{R} \partial_{R}
		- \frac{q^2}{4} B^2 R^2
	\right] X = - \Lambda^2 X,\\
	\partial_{r}^2 F
	+ \left(
		\frac{f'}{f} + \frac{h'}{h}
	\right) \partial_{r} F
	+ \frac{f^{-1} q^2 A_0 - M^2}{f}F
	- \frac{\Lambda^2}{f h} F = 0.
\end{gather}
where $\Lambda$ is a constant eigenvalue.
Rewriting $R = s /\sqrt{q B}$, we have
\begin{equation}
	\partial_{s}^2 X + \frac{1}{s} \partial_{s} X
	- \frac{1}{4} s^2 X
	= -   \lambda^2 X,\quad
	\lambda^2=\frac{\Lambda^2}{q B}.
\end{equation}
Imposing $X(s=\infty)=0$, we obtain the solution as

\begin{equation}
	X = X_{0} e^{s^2/4} L_{-\frac{1+\lambda^2}{2}}\left(-\frac{s^2}{2}\right),\quad
    \lambda
    = \sqrt{2 n + 1}, \quad (n=0,1,2\cdots),
\end{equation}
where $X_{0}$ is a normalization constant, and $L_{m}(x)$ denotes Laguerre polynomials.
The higher modes are oscillating, whereas the lowest mode is given by a simple Gaussian function:

\begin{equation}
	X = X_0 e^{-s^2/4},\quad \Lambda = \sqrt{q B}.
\end{equation}%
Using this solution, the equation for $F$ becomes
\begin{equation}
	\partial_{r}^2 F
	+ \left(
		\frac{f'}{f} + \frac{h'}{h}
	\right) \partial_{r} F
	+ \frac{f^{-1} q^2 A_0 - M^2}{f}F
	- \frac{q B}{f h} F = 0.
\end{equation}
We can investigate the onset of the instability by solving the above ODE with an appropriate boundary condition.
Setting $M^2 = -2$ corresponding to the scaling dimensions $\Delta = 1,2$, we obtain the asymptotic expansion as
\begin{equation}
    F(r) = r^{-1} F_{(1)} + r^{-2} F_{(2)} + \order{r^{-3}}.
\end{equation}
In the standard quantization, $F_{(1)}$ can be understood as a source term.
We impose a vanishing Dirichlet condition, i.e., $F_{(1)}=0$, to find the onset of the instability.

\begin{figure}[htbp]
    \centering
    \includegraphics[width=7cm]{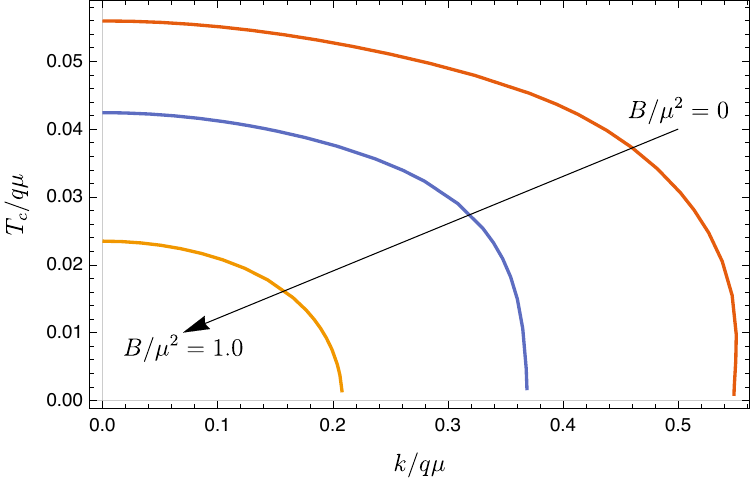}
    \includegraphics[width=7cm]{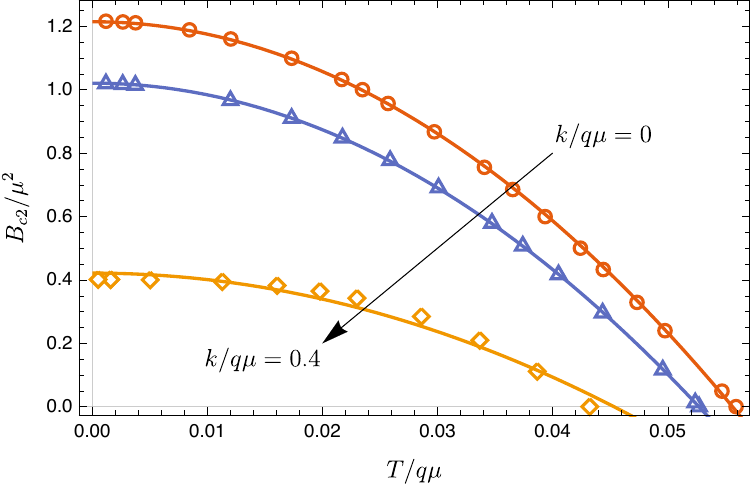}
    \caption{
    Onset of the charged scalar instabilities.
    We set $q=6$ and $M^2=-2$.
    (left) In the $T$--$k$ plane for fixed $B/\mu^2$.
    The curves correspond to $B/\mu^2 =0,0.5,1.0$ from outside to inside.
    (right) In the $B$--$T$ plane for fixed $k/q\mu$.
    The data correspond to $k/q\mu = 0, 0.2, 0.4$ from outside to inside.
    The curves show fitting results with eq.~(\ref{eq:fit_Bc2}), while the points show the numerical data.
    Note that we wrote $T$ as $T_{c}$ in the left panel, and $B$ as $B_{c2}$ in the right panel.
    }
    \label{fig:phase_boundary}
\end{figure}

The phase boundaries are obtained as figure \ref{fig:phase_boundary}.
We set $q=6$.
We wrote $T$ and $B$ of the phase boundaries as $T_{c}$ and $B_{c2}$, as appropriate.
As it is shown in the left panel, the phase boundary shrinks as $B$ increases, i.e., the ordered phase is suppressed by applying magnetic field.
As it is shown in the right panel, the critical magnetic field $B_{c2}(T)/\mu^2$ shows a typical $1-T^2/T_{c}^2$ behavior for each value of $k/q\mu$.
The curves denote fitting results by
\begin{equation}\label{eq:fit_Bc2}
    B_{c2}(T)= B_{c2}^{T=0} \left[
        1 - \left(\frac{T}{T_{c}^{B=0}}\right)^{2}
    \right],
\end{equation}
where $B_{c2}^{T=0}$ and $T_{c}^{B=0}$ are fitting parameters.
The numerical data can be fitted well by the above function when $k$ is small.

\section{Comment on ST-type Gubser-Rocha model}\label{STGR}
In this section, we leave a comment on the ST-type Gubser-Rocha model proposed in \cite{Xu:2023qlu}.
If one reads the mass of the axion correctly, it violates the BF bound in this model.
The Lagrangian density of the ST-type Gubser-Rocha model is given by
\begin{equation}
	\frac{\mathcal{L}_{\mathrm{ST}}}{\sqrt{-g}} =
	R - \frac{3}{2} \frac{\partial_{\mu} \tau \partial^{\mu} \bar{\tau}}{(\Im \tau)^2}
	- \frac{1}{4} e^{-\phi} F^2 + \frac{1}{4} \chi F \tilde{F}
	+ \frac{3}{L^2} \frac{\tau\bar{\tau} + \Re\tau + 1}{\Im\tau}.
\end{equation}
This is the ``ST-completion" of the Gubser-Rocha model.
Expanding $\tau$, we can write
\begin{equation}
	\frac{\mathcal{L}_{\mathrm{ST}}}{\sqrt{-g}} =
	R - \frac{3}{2} (\partial \phi)^2 - \frac{3}{2}e^{2\phi}(\partial\chi)^2 
	- \frac{1}{4} e^{-\phi} F^2 + \frac{1}{4} \chi F \tilde{F}
	+ \frac{1}{L^2}\left(
		6 \cosh\phi + 3 (\chi^2 + \chi) e^{\phi}
	\right).
\end{equation}
With the linear axions term $\mathcal{L}_{2}$, the following analytic solution is available:
\begin{equation}
\begin{gathered}
	f(r) = \sqrt{r(r+\rs)^3}\left(
		\frac{\sqrt{3}}{2 L^2} - \frac{k^2}{2(r+\rs)^2}
		- \frac{\sqrt{3}(\mu^2 + P^2)(r_0 + \rs)^2}{6 \rs (r+\rs)^3}
	\right),\quad
	h(r) = \sqrt{r(r+\rs)^3},\\
	e^{\phi} = 2 \frac{\mu^2 r + P^2(r+\rs)}{(P^2 + n^2)\sqrt{3 r (r+\rs)}},\quad
	\chi = - \frac{\mu^2 r + \sqrt{3}P \mu \rs + P^2(r+\rs)}{2(\mu^2 r + P^2 (r+\rs))},\\
	A = \mu\left(
		1 - \frac{r_0 + \rs}{r+\rs}
	\right)\dd{t} - P (r_0 + \rs) y \dd{x},\quad
	\psi_{I} = k x^{I},
\end{gathered}
\end{equation}
where
\begin{equation}
	\mu = \left[
		2\sqrt{3}\rs(r_0 + \rs)\left(
			\frac{\sqrt{3}}{2} - \frac{k^2}{2 (r_0 + \rs)^2} - \frac{P^2}{2\sqrt{3}\rs (r_0 + \rs)}
		\right)
	\right]^{1/2}.
\end{equation}
The charge density
\begin{equation}
    n = \frac{1}{2} (P + \sqrt{3}\mu)(r_0 + \rs),
\end{equation}
is shifted by the magnetic field, which is a demonstration of the Witten effect.
The magnetic field is $B = P(r_0 + \rs)$.
The asymptotic behavior of the metric is written as
\begin{equation}
\begin{aligned}
	\dd{s}^2 \sim -\frac{\sqrt{3}}{2 L^2} r^2 \dd{t}^2  + r^2 \dd{x}^2
	+ \frac{\sqrt{3} L^2}{2} \frac{\dd{r}^2}{r^2}
	=
	- \frac{r^2}{l^2} \dd{t}^2  + r^2 \dd{x}^2
	+ l^2 \frac{\dd{r}^2}{r^2},
\end{aligned}
\end{equation}
where $l^2=2L^2/\sqrt{3}$ corresponding to the AdS radius.
The temperature can be computed as
\begin{equation}
	T = \frac{1}{16\pi \sqrt{r_0(r_0 + \rs)^3}}\left(
		-2 k^2 r_0 + (r_0 + \rs)^2\left(
			2 (4 + \sqrt{3}) r_0 - (\sqrt{3}-2) \rs
		\right)
	\right).
\end{equation}
where we have set $l=1$.

Let us now consider the bulk masses of the dilaton and the axion fields.
The potential term is expanded as
\begin{equation}
	6 \cosh\phi + 3 (\chi^2 + \chi) e^{\phi}
	=
	6 + 3 \chi + 3 (\phi^2 + \phi \chi + \chi^2) + \order{\phi,\chi}^3.
\end{equation}
As it is, the bulk mass is unclear.
Since the potential has a local minimum at $\chi=-1/2$ and $\phi = \log\frac{2}{\sqrt{3}}$, we should expand the fields around this point.
Now, we redefine the axion and the dilaton as $\chi = -1/2 + \hat{\chi}$ and $\phi = \log\frac{2}{\sqrt{3}} + \hat{\phi}$.
In terms of the new fields, the potential term is expanded as
\begin{equation}
	6 \cosh\phi + 3 (\chi^2 + \chi) e^{\phi}
	=
	3\sqrt{3} + \frac{3\sqrt{3}}{2} \hat{\phi}^2 + 2\sqrt{3}\hat{\chi}^2 + \order{\hat{\phi},\hat{\chi}}^3.
\end{equation}
We can read off the masses as
\begin{equation}
	m_{\hat{\phi}}^2 = - \frac{\sqrt{3}}{L^2},\quad
	m_{\hat{\chi}}^2 = - \frac{4}{\sqrt{3}L^2}.
\end{equation}
Note that the normalization is now $e^{-1}\mathcal{L} = - \frac{3}{2}(\partial \varphi)^2 - \frac{3}{2}m^2 \varphi^2$.
As we have mentioned, $L$ is not the AdS radius because the cosmological constant term is given by $3\sqrt{3}/L^2$.
Using $l^2 = 2 L^2/\sqrt{3}$, we obtain
\begin{equation}
	m_{\hat{\phi}}^2 = - \frac{2}{l^2},\quad
	m_{\hat{\chi}}^2 = - \frac{8}{3l^2}.
\end{equation}
The dilaton mass is unchanged from one in the S-completion model.
On the other hand, the axion mass is changed, and $m^2l^2 = - 8/3 \approx - 2.67$ violates the BF bound $-9/4= -2.25 \leq m^2l^2$.
The violation of the BF bound implies instability in the dual QFT, so this model would not be suitable as a holographic model.

\bibliography{main}
\bibliographystyle{JHEP}
\end{document}